\documentclass[aps,prb,twocolumn,floatfix,footinbib]{revtex4}
\usepackage{amssymb}
\usepackage{graphicx}%
\usepackage{epstopdf}
\begin{document}
\title{Silicon spin communication} 
\author{Hanan Dery,$^{1,2}$  Yang Song,$^{2}$ Pengke Li,$^{1}$
and Igor \v{Z}uti\'{c}$^{3}$}
\affiliation{
$^{1}$ Department of Electrical and Computer Engineering,
University of Rochester,
Rochester, New York 14267, USA \\
$^{2}$ Department of Physics and Astronomy, University of Rochester,
Rochester, New York 14267, USA \\
$^{3}$ Department of Physics, University at Buffalo, State University of
New York, New York 14260, USA}

\begin{abstract}
Recent experimental breakthroughs have demonstrated that the electron spin
in silicon can be reliably injected and detected as well as transferred over
distances exceeding 1~mm. We propose an on-chip communication paradigm which
is based on  modulating spin polarization of a constant current in silicon
wires. We provide figures of merit for this scheme by studying spin relaxation
and drift-diffusion models in silicon.
\end{abstract}
\maketitle

Virtually all modern information technologies are based on modulating
electromagnetic waves whose propagation is described by Maxwell equations.
When this technique is used for on-chip communication, signals are transmitted
via metallic wires, modeled as transmission lines with the voltage and current
being distance and time-dependent. The main drawbacks of this technique relate
to dynamical crosstalk between wires, RC
bottlenecks, and electromigration between silicon and its interconnect material.\cite{ITRS,Deutsch1998:PIEEE} These effects become increasingly
acute with reducing the spacing between adjacent wires and with increasing
the modulation frequency. Implementing new high-performance communication
schemes and interconnects is thus central to the scaling of integrated circuit
technologies.

We propose a different concept for data communications which relies on the modulation of the
electrons'  spin polarization of a \textit{constant current} in Si wires.
If spin, rather than voltage, encodes information, then the wires remain
charged indefinitely where the constant charge current is used to drive the
information but not to carry it. This scheme is free of dynamical transmission
line effects, electromigration problems and the need for wire shielding. Using
the electron spin to process\cite{Dery2007:N,Dery2011:IEEE} or transfer
information in semiconductors may spur spintronic applications beyond
information storage.\cite{Zutic2004:RMP,Maekaea_Book}

Silicon is a promising material candidate due to its weak spin-orbit
coupling.\cite{Zutic2006:PRL,Cheng2010:PRL,Li2010:PRL,Li2011:P} Measured spin lifetime of intrinsic Si at low temperatures is within the $\mu$s range,\cite{Huang_PRL07} and 10 ns at 300~K,\cite{Lancaster_PPSL64,Lepine_PRB70}
the longest of any inorganic bulk semiconductor. At the saturation drift
velocity of silicon ($\sim$10$^7$~cm/s), this corresponds to a transport
length scale exceeding 1~mm. Furthermore, recent advances in spin injection into Si~\cite{Appelbaum2007:N,Jonker2007:NP,Dash2009:N,Ando2010:APL,Suzuki2011:APE,Li_nature_com11}
are encouraging for experimental implementation of spin interconnects. Figure~1 shows a scheme of the spin-based communication protocol.
The transmitter generates information via modulation of the magnetization in
the upper spin-injector contact. This is realized by modulating the voltage
of a short and local transmission line above the contact. The current, $I_0$,
across the ferromagnet/Si interface is constant and driven by an external
source. $I_0$ is comprised primarily of electrons whose spin matches the
spin-up population of the injector contact. This constant current flows
in the Si wire without interfering with the information that may propagate in
the adjacent wires (not shown). At the receiver, $I_0$ is split into two paths where electrons
prefer to be extracted from a ferromagnetic contact whose magnetization
direction matches their own spin direction. Thus, the current in one of the
receiver's contacts is greater than in the other contact and a local
differential device/amplifier may resolve the encoded information.

\begin{figure}
\includegraphics[scale=0.21]{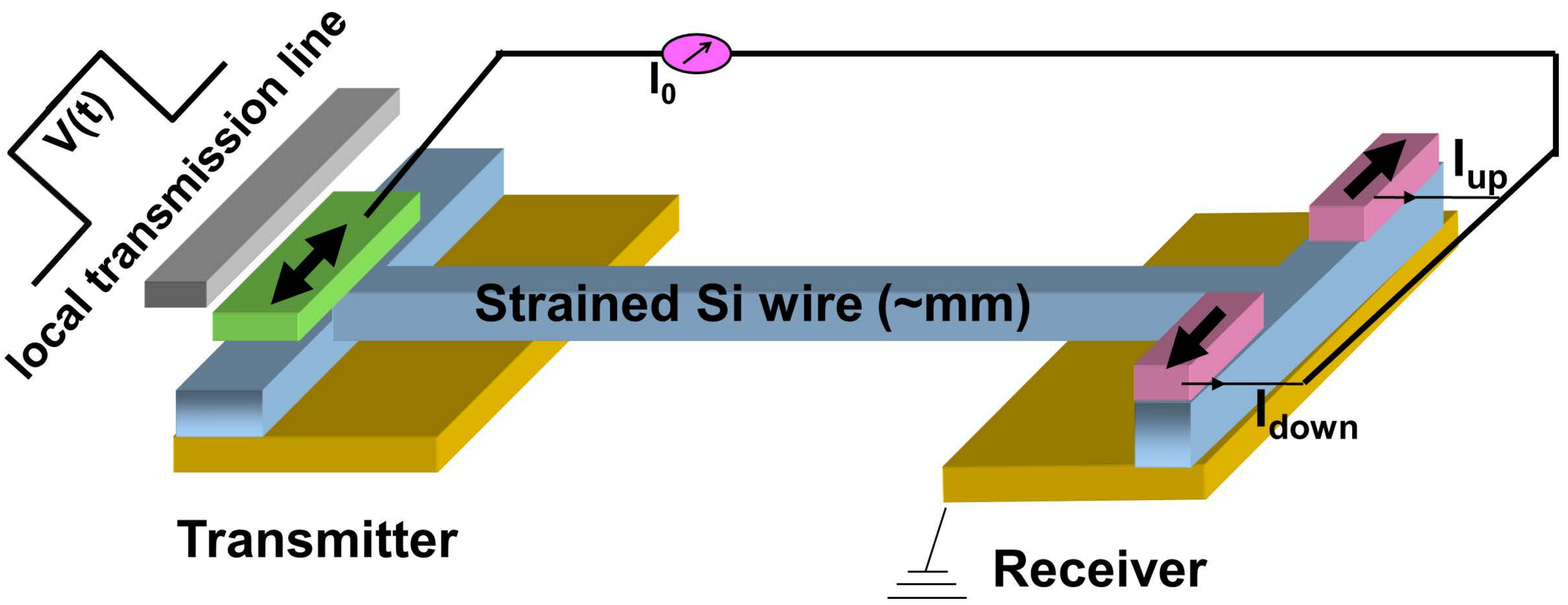}
\caption{(Color online) A spin-based communication scheme. The current in
Si wire $I_0$, is constant, but it is mainly governed by a spin-up
or a spin-down current depending on the direction of the injector
magnetization (left). The receiver splits the current into two paths
(right contacts.) The detection is ``1'' or ``0'' if the current is governed,
respectively, by spin-up ($I_\mathrm{up} >I_\mathrm{down}$) or
spin-down ($I_\mathrm{up} < I_\mathrm{down}$), currents.}
\label{fig:Scheme}
\end{figure}
The spin relaxation time $\tau_s$  and mobility $\mu$ 
in the Si wire are the most
important parameters in setting the proposed on-chip communication
lengthscale. We consider non-degenerate Si wires
with cross section areas
larger than 10$\times$10~nm$^2$. Since the effective electron Bohr radius and
the mean free path of thermal electrons in Si at 300~K is of the order of
a few nm, the transport in such wires is bulk-like and the spin relaxation is
governed by Elliott-Yafet processes. \cite{Elliott_PR54,Yafet_1963}
Taking into account the six conduction band valleys in silicon, we follow the
classical categorization into intravalley and intervalley scattering where the
latter has contributions from $g$ and $f$ processes which denote,
respectively, scattering between opposite valleys and between valleys on
different crystal axes.\cite{Cardona_Book}
\begin{figure}
\includegraphics[width=8cm, height=4.5cm]{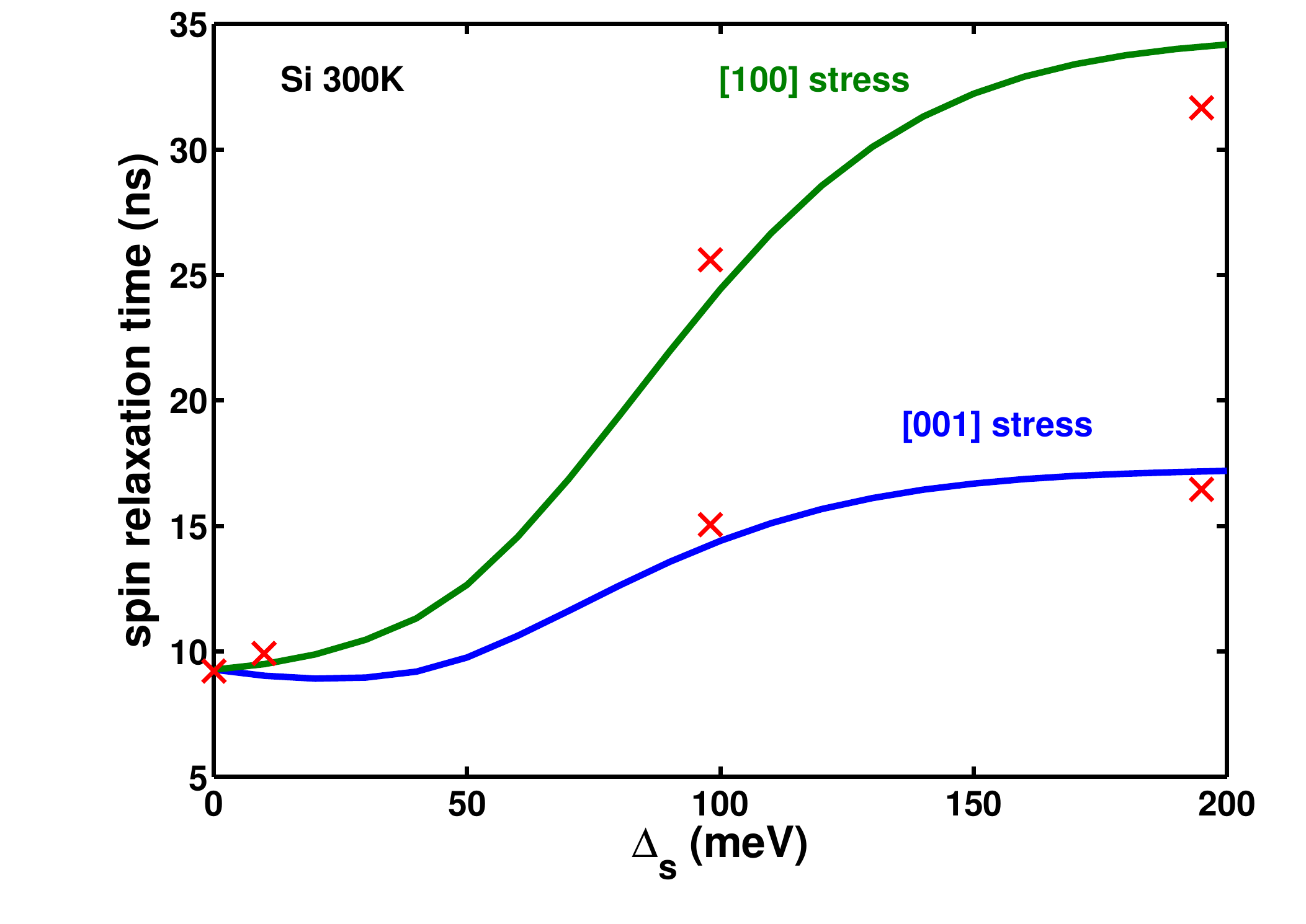}
\caption{(Color online) Calculated spin relaxation time as a function of
the valley splitting, $\Delta_s$,  due to stress along the [001] and [100]
crystal axes. The spin quantization axis is chosen parallel to the [001].
The simulated stress levels are compatible with external stress levels in bulk
silicon or with internal strain in heterostructures with slight lattice mismatch.\cite{hetero}
X symbols denote elaborate numerical results and solid lines are calculated
via weighted suppression of the numerical contributions of the unstrained
case ($\Delta_s$=0).}
\label{fig:relax}
\end{figure}

We first study how $\tau_s$ 
can be tailored using strain
and then we incorporate our findings into a spin dependent drift-diffusion
description of the wire.\cite{Zutic2004:RMP,Maekaea_Book}
%
The strain has two effects
which increase $\tau_s$. 
The first 
is to lift the energy degeneracy between valleys that reside on different
axes (valley splitting) which suppresses
the $f$-process contribution to spin-relaxation.
\cite{Tang2011:P, Cheng2010:PRL} In an unstrained Si this contribution
is dominant at high-temperatures.\cite{Li2011:P} Valley splitting is possible
if the diagonal strain components are not equal (e.g., via [001] and [011]
stress configurations). The second effect is to increase the energy spacing
between the bottom of the conduction band and the spin hot-spot at the edge of
the Brillouin zone.\cite{Li2011:P, Cheng2010:PRL} Spin hot-spots give rise to
fast
spin relaxation due to the enhanced spin-mixing of states in these
regions.\cite{Fabian_PRL98} Application of strain with nonzero off-diagonal
components (e.g., via [111] and [011] stress configurations) can transfer the
two-band degeneracy from the X-point region to farther energy regions in the
Brillouin zone. 
A comprehensive examination of the strain effects will be studied elsewhere. Here we summarize the results
that are most relevant to the spin communication protocol.

Figure~2 shows simulation for 
the spin relaxation suppression in Si as a function of stress levels along
the [001] and [100] crystal axes. These simulations follow a numerical
procedure similar to the unstrained case.\cite{Cheng2010:PRL} Strain effects
were incorporated into the band structure calculation following
Ref.~[\onlinecite{Rieger_PRB}] and into the phonon dispersion and polarization
vectors following Ref.~[\onlinecite{Eyrigit_PRB96}]. The figure shows
that $\tau_s$ 
at 300 K 
can be four times longer than
the unstrained case. This optimal improvement is reached when the spin
quantization axis (chosen parallel to the [001] crystal axis) is perpendicular
to the axis of valleys with lower energy. The saturation of $\tau_s$ 
at higher stress levels is reached when $f$-processes are effectively quenched
and the dominant contribution to spin relaxation comes from intravalley
scattering,
\begin{eqnarray}
\frac{\tau_{s}(\Delta_s \gg k_BT)}{\tau_{s}(\Delta_s = 0)} = \frac{4}{3}\frac{1}{1+\delta_{\hat{e},\hat{z}}} \left( 1 + \frac{ \tau_{i,0}\tau_{g,0} }{\tau_{f,0}(\tau_{i,0} + \tau_{g,0})  } \right)  \,\,, \label{eq:improve_factor}
\end{eqnarray}
where $\hat{e}$ and $\hat{z}$ are unit vectors along the stress and spin quantization axes, respectively. $\tau_{i,0}$, $\tau_{g,0}$ and $\tau_{f,0}$ denote, respectively, components of spin relaxation in the unstrained case due to intravalley, $g$-process and $f$-process scattering. Their analytical forms are given in Ref.~\onlinecite{Li2011:P}.

Our findings are considerably more conservative than those of
Tang \textit{et al.} who predicted an order of magnitude longer
$\tau_s$ 
when the $f$-processes are  quenched.\cite{Tang2011:P}
The reason for the discrepancy is that in our case the intravalley
rate due to scattering with acoustic phonons is much faster,
$\tau_{i,0} \sim$10~$\mu$s at 50~K and $\tau_{i,0} < 50$~ns at 300~K,
consistent with Larmor precession and spin-valve measurements of spin injection
in Si,\cite{Huang_PRL07,Appelbaum2007:N,Huang_PRB10} with
previous detailed numerical calculation,\cite{Cheng2010:PRL} and with
analytical $k\cdot p$ formalism.\cite{Li2011:P} It is possible that the
coupling between the upper and lower conduction bands which provides the
dominant contribution to intravalley spin relaxation,\cite{Li2011:P} is
not included in the $sp^3$ model in Ref.~[\onlinecite{Tang2011:P}].

\begin{figure}
\includegraphics[width=8.5cm]{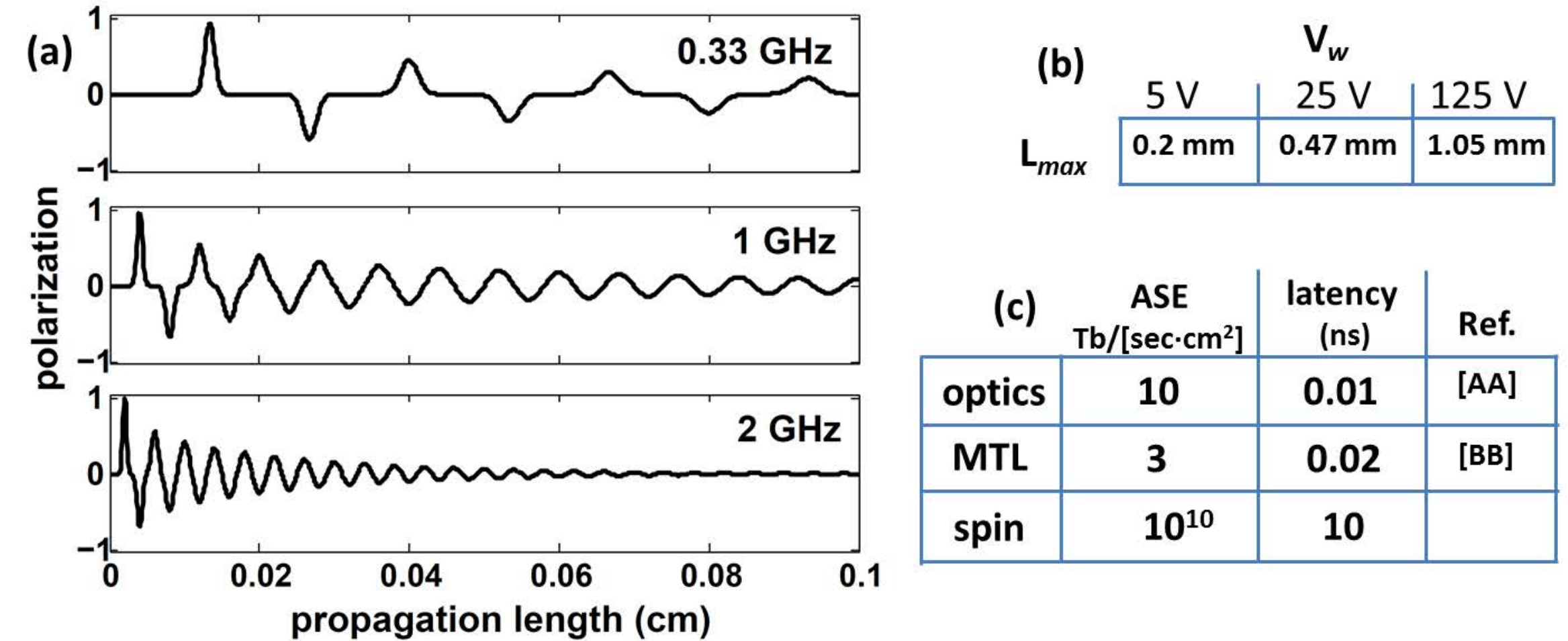}
\caption{(a) Current polarization of an alternating pulse train (spin-up, spin-down, spin-up ...) along a strained Si wire at three repetition rates. (b) Maximal propagation for three voltage drop levels across the wire. (c) Top view scheme of a spin interconnect circuit. At 50\% areal filling with wires whose geometry and conductivity parameters are shown on the right hand side, the bandwidth is
1300 Tbit/(s$\cdot$ cm$^2$) for 5~V bias and 0.13~GHz spin pulse modulation.}
\label{fig:train}
\end{figure}

To explore the feasibility of the proposed communication scheme,
we model spin-dependent pulse propagation in a strained Si wire.
Using a drift-diffusion model, the propagation of a pulse is, 
\begin{eqnarray}
P(x,t) = \frac{P_0}{\sqrt{4\pi D t}} \exp\left( \frac{(x- \mu E t)^2}{4 D t}
- \frac{t}{\tau_s}  \right)   \,\,, \label{eq:f_mat}
\end{eqnarray}
where $P(x,t)$ denotes the current spin polarization at location $x$ and time
$t$,  $P_0$ is the polarization at the transmitter,
$D$ is the diffusion constant, and $E$ is the electric field.
Using the room temperature parameters of strained Si, $\tau_s$=35~ns from
Fig.~\ref{fig:relax} and
$\mu$=2500 cm$^2$/V$\cdot$s,\cite{Ungersboeck_IEEE07} we model propagation of
a (spin) pulse train across a strained Si wire. Figure~\ref{fig:train}(a) shows a snapshot of the current polarity along the 1~mm wire
due to transmission of a pulse train at three different repetition rates. The drift velocity is $\mu$$E$=4$\cdot$10$^6$~cm/sec. We consider the
`worst case scenario' in which the polarity is flipped with each pulse. After 1~mm propagation the current remains spin-polarized at 1~GHz pulse repetition rate where the peak polarity drops by an order of magnitude (from $\pm$$P_0$ to $\pm$0.1$P_0$).

To provide figures of merit 
of this technique, we define the maximal repetition rate, $f_{max}$, which is mostly dictated by the intermixing of nearby spin-pulses (diffusion effect). If we limit the intermixing to 10\% then,
\begin{eqnarray}
f_{max} \!=\! \frac{1}{3}\sqrt{\!\frac{qV^3}{k_BT}} \frac{\mu}{L^2}\,,\,\,\,\, \tau_{\ell}\!=\!\sqrt{\!\frac{L^2\tau_s}{\mu V}}\,,\,\,\,\, L_{max} \!= \! \sqrt{\!\mu V \tau_s}\,,\,\,\, \label{eq:merits}
\end{eqnarray}
where $V$ is the voltage drop across the wire and $k_BT/q$ is the thermal voltage,
 $\tau_{\ell}$ is the latency (propagation delay) and $L_{max}$ is the maximal
propagation length limited by spin relaxation ($L < L_{max}$).
Using the previous 
$\tau_s$ and $\mu$ parameters, the table in Fig.~\ref{fig:train}(b) shows the maximal
propagation lengths of the communication scheme. Figure~3(c) shows a scheme
of a lateral circuit whose interconnects are strained silicon wires with the
parameters on the right hand side of the figure. Assuming a 50\% areal filling with wires,
the attainable bandwidth for such a lateral circuit working at 5~V and  0.13~GHz
 is 1300~Tbit/(s$\cdot$cm$^2$) 
and the needed power is 0.25~Watt (Joule heating of the wires).
These bandwidth and power are by far superior to 
any other existing technique (see, e.g., Fig.~3 in Ref.~[\onlinecite{McFadden_AO06}]). On the other hand, the main drawback of the proposed communication scheme is the increased latency ($>$10~ns per 1~mm compared with 10-20~ps in metallic transmission lines and optics on-chip). Nonetheless, the extremely high bandwidth of spin interconnects may satisfy wiring demands between nearest neighbor nodes of practical multi-core microprocessor architectures and they may also spur communication schemes in emerging 3D electronic technologies, reconfigurable logic and optoelectronic systems (e.g.; in spin-lasers by modulating the information via the polarization rather than the amplitude of light\cite{Lee2010:APL,Holub2007:PRL,Rudolph2003:APL}).

In conclusion, we have studied the performance of spin interconnects implemented in silicon wires. By using strain, the propagation distances are predicted to reach the 1~mm scale at room temperature while keeping a high fidelity of the signal, demonstrating the feasibility of high-performance spin interconnects. We have also provided figures of merit 
for the maximal spin-pulse repetition rate and propagation length as well as
discussing the latency versus bandwidth trade-offs.



This work was supported by the AFOSR-DCT FA9550-09-1-0493, NSF-ECCS 0824075,
NSF-ECCS CAREER 054782, and U.S. ONR N0000140610123.




\begin{thebibliography}{99}

\bibitem{ITRS}
2009 International Technology Roadmap for Semiconductors www.itrs.net

\bibitem{Deutsch1998:PIEEE}
A. D. Deutsch,
Proc. IEEE {\bf 86}, 315 (1998).

\bibitem{Dery2007:N}
H. Dery, P. Dalal, L. Cywinski, and L.~J. Sham, Nature {\bf 447}, 573 (2007);

\bibitem{Dery2011:IEEE}
H. Dery, H. Wu, B. Ciftcioglu, M. Huang, Y. Song, R. Kawakami, J. Shi,
I. Krivorotov, I. \v{Z}uti\'c, L.~J. Sham, arXiv:1101.1497, preprint.

\bibitem{Zutic2004:RMP}
I. \v{Z}uti\'c, J. Fabian, and S. Das Sarma,  Rev. Mod. Phys. {\bf 76},
323 (2004); J. Fabian, A. Mathos-Abiague, C. Ertler, P. Stano, and
I. \v{Z}uti\'c,  Acta Phys. Slov. {\bf 57}, 565 (2007).

\bibitem{Maekaea_Book} S. Maekawa, \textit{Concepts in Spin Electronics}, (Oxford, New-York, 2006).

\bibitem{Zutic2006:PRL}
I. \v{Z}uti\'c, J. Fabian, and S.~C. Erwin,  Phys. Rev. Lett. {\bf 97},
026602 (2006).

\bibitem{Cheng2010:PRL}
J.~L. Cheng, M.~W. Wu and J. Fabian, Phys. Rev. Lett. {\bf 104}, 016601 (2010).

\bibitem{Li2010:PRL}
P. Li and H. Dery, Phys. Rev. Lett. {\bf 105}, 037204 (2010).

\bibitem{Li2011:P}
P. Li and H. Dery, arXiv:1103.3800, preprint.

\bibitem{Huang_PRL07} B. Huang, D. J. Monsma, and I. Appelbaum, Phys. Rev. Lett. \textbf{99}, 177209 (2007); B. Huang, H.-J. Jang, and I. Appelbaum, Appl. Phys. Lett. \textbf{93}, 162508 (2008);

\bibitem{Lancaster_PPSL64} G. Lancaster, J. A. Van Wyk, and E. E. Schneider, Proc. Phys. Soc. London \textbf{84}, 19 (1964).

\bibitem{Lepine_PRB70} D. J. L\'{e}pine, Phys. Rev. B \textbf{2}, 2429 (1970).

\bibitem{Appelbaum2007:N}
I. Appelbaum, B. Q. Huang, and D. J. Monsma, Nature {\bf 447}, 295 (2007).

\bibitem{Jonker2007:NP}
B.~T. Jonker, G. Kioseoglou, A.~T. Hanbicki, C.~H. Li, and P.~E. Thompson,
Nature Phys. {\bf 3}, 542 (2007).

\bibitem{Dash2009:N}
S.~P. Dash, S. Sharma, R.~S. Patel, M.~P. de Jong, and R. Jansen, Nature
{\bf 462}, 491 (2009).

\bibitem{Ando2010:APL}
Y. Ando, K. Hamaya, K. Kasahara, Y. Kishi, K. Ueda, K. Sawano, T. Sadoh, and M. Miya, Appl.  Phys.  Lett. {\bf 94}, 182105 (2009). 

\bibitem{Suzuki2011:APE}
T. Suzuki, T. Sasaki, T. Oikawa, M. Shiraishi, Y. Suzuki, and K. Noguchi,
App. Phys. Express {\bf 4}, 023003 (2011).

\bibitem{Li_nature_com11}
C. H. Li, O. M. J. van 't Erve, and B. T. Jonker, Nature Communications \textbf{2}, 245 (2011). 

\bibitem{Elliott_PR54} R. J. Elliott, Phys. Rev. \textbf{96}, 266 (1954).

\bibitem{Yafet_1963} Y. Yafet, in \textit{Solid State Physics}, edited by F. Seitz and D. Turnbull (Academic, New York, 1963), Vol. 14, p.~1.

\bibitem{Cardona_Book} P. Y. Yu and M. Cardona, \textit{Fundamentals of Semiconductors} (Springer, Berlin, 2005), 3$^{rd}$ Ed., Ch.~2-5.

\bibitem{Tang2011:P} J.-M. Tang, B. T. Collins, and M. E. Flatt\'{e}, arXiv:1104.9705, preprint.


\bibitem{Fabian_PRL98} J. Fabian and S. Das Sarma, Phys. Rev. Lett. \textbf{81}, 5624 (1998).


\bibitem{Rieger_PRB} M. M. Rieger and P. Vogl, Phys. Rev. B \textbf{48}, 14276 (1993).

\bibitem{Eyrigit_PRB96}  by R. Eyri\u{g}it and I. R. Herman, Phys. Rev. B \textbf{53}, 7775 (1996). 

\bibitem{hetero} Heterostructures with large internal strain are undesirable since the stressed layer cannot be grown beyond a few nm without strain relaxation (e.g. in Si$_{1-x}$Ge$_x$/Si with $x>0.15$). Accordingly, in such confined structures one should also consider spin relaxation due to interfacial scattering and structural inversion asymmetry.


\bibitem{Huang_PRB10} B. Huang and I. Appelbaum, Phys. Rev. B 82, 241202(R) (2010).
%

\bibitem{Ungersboeck_IEEE07} E. Ungersboeck, S. Dhar, G. Karlowatz,
V. Sverdlov, H. Kosina, and S. Selberherr, IEEE Trans. Electron
Device \textbf{54}, 2183 (2007). 

\bibitem{McFadden_AO06} M. J. McFadden, M. Iqbal, T. Dillon, R. Nair, T. Gu, D. W. Prather, and M. W. Haney, Appl. Opt. \textbf{45}, 6358 (2006). 

\bibitem{Lee2010:APL}
J. Lee, W. Falls, R.  Oszwa\l dowski, and I. \v{Z}uti\'c,
Appl.  Phys.  Lett. {\bf 97}, 041116 (2010).

\bibitem{Holub2007:PRL}
M. Holub, J. Shin, D. Saha, and P. Bhattacharya,
Phys. Rev. Lett. {\bf 98}, 146603 (2007).

\bibitem{Rudolph2003:APL}
J. Rudolph, D.~H\"{a}gele, H.~M. Gibbs, G. Khitrova,
and M. Oestreich,  Appl. Phys. Lett. {\bf 82}, 4516 (2003).

\end{thebibliography}
\end{document}